**Correspondence of Bell State and One-Particle State Transformations**

R. G. Beil[1,2]

_______________________________________________________________

There is a direct correspondence between two-particle, entangled quantum states, for example, Bell states, and the relative values of the component one-particle states. This leads to a new rationale for quantum computing which makes use of sequential processing of one-particle states rather than the parallel processing associated with multiparticle states. It is shown that deterministic transformations can correspond to certain Bell state operations. There are some implications for the continuing discussion of quantum realism and entanglement. A principle of *relative realism* is advocated

_______________________________________________________________

KEY WORDS: entanglement, Bell states, quantum computing

PACS: 03.65.Bz; 03.67.Lx; 42.50.Dv

[1]313 S. Washington, Marshall, Texas 75670; e-mail; rbeil@etbu.edu

[2]Institute for Studies in Pragmaticism, Texas Tech University, Lubbock, Texas, 79409-0002

We first review a simple and well-known formalism for the study of quantum computing. This will establish notation and provide a basis for further developments. Useful sources for this background are the book of Nielsen and Chuang (Nielsen and Chuang, 2000) and the papers of Bruss and Gudder (Bruss, 2002; Gudder, 2003).

We assume that a single particle has orthonormal pure states $|0\rangle$ and $|1\rangle$. A mixed state is

$$|a\rangle = a_0 |0\rangle + a_1 |1\rangle \qquad (1)$$

The development here assumes only one degree of freedom for a single particle. The case of several degrees of freedom remains to be given.

A one-particle operator is

$$T = T_{00} |0\rangle\langle 0| + T_{01} |0\rangle\langle 1| + T_{10} |1\rangle\langle 0| + T_{11} |1\rangle\langle 1| \qquad (2)$$

Examples are the identity $I = |0\rangle\langle 0| + |1\rangle\langle 1|$ and the "flip" $F = |0\rangle\langle 1| + |1\rangle\langle 0|$. The notation for the matrix element is $T_{ij} = \langle i | T | j \rangle$. (We recognize that there are limits to the experimental realization of operations such as the flip. We assume that some optimal approximations to these operations exist (De Martini et al, 2002; Fiorentino and Wong, 2004)).

We will frequently use a convenient matrix notation established by the examples,

$$|a\rangle = \begin{bmatrix} a_0 \\ a_1 \end{bmatrix}, \qquad T = \begin{bmatrix} T_{00} & T_{01} \\ T_{10} & T_{11} \end{bmatrix} \qquad (3)$$

The representation of the action of a general operator on a state is,

$$\begin{bmatrix} a_0{}' \\ a_1{}' \end{bmatrix} = \begin{bmatrix} T_{00} & T_{01} \\ T_{10} & T_{11} \end{bmatrix} \begin{bmatrix} a_0 \\ a_1 \end{bmatrix} \qquad (4)$$

For example, for the transformations

$$T_{\pm} = \begin{bmatrix} 1 & 0 \\ 0 & \pm 1 \end{bmatrix} \qquad (5)$$

one obtains another orthogonal basis

$$|+\rangle = a_0|0\rangle + a_1|1\rangle$$

$$|-\rangle = a_0|0\rangle - a_1|1\rangle \qquad (6)$$

which is normalized by the condition

$$a_1 = \left(1 - a_0{}^2\right)^{\frac{1}{2}} \qquad (7)$$

The usual special case is $a_0 = a_1 = 1/\sqrt{2}$.

In general, however, the two single-particle states depend on a single parameter, say, $a_0$,

which has values over a continuous range ($0 \leq a_0 \leq 1$).  This means that single-particle measurements are basically probabilistic.

Two-particle states formed from two of these one-particle states can be derived from the

tensor product

$$|ab\rangle = |a\rangle \otimes |b\rangle = \begin{bmatrix} a_0 \begin{bmatrix} b_0 \\ b_1 \end{bmatrix} \\ a_1 \begin{bmatrix} b_0 \\ b_1 \end{bmatrix} \end{bmatrix} = \begin{bmatrix} a_0 b_0 \\ a_0 b_1 \\ a_1 b_0 \\ a_1 b_1 \end{bmatrix} \qquad (8)$$

This product is not, in general, commutative.  Those two-particle states which can be

written as a tensor product are said to be *separable.*

For the pure one-particle component states, the two-particle states are

$$|00\rangle = \begin{bmatrix} 1 \\ 0 \\ 0 \\ 0 \end{bmatrix}, \qquad |01\rangle = \begin{bmatrix} 0 \\ 1 \\ 0 \\ 0 \end{bmatrix}, \qquad |10\rangle = \begin{bmatrix} 0 \\ 0 \\ 1 \\ 0 \end{bmatrix}, \qquad |11\rangle = \begin{bmatrix} 0 \\ 0 \\ 0 \\ 1 \end{bmatrix} \qquad (9)$$

A new rationale for these states has recently been given (Beil and Ketner, 2003; Beil,

2004).

These two-particle component states are obviously orthogonal.   They are called the computational basis of the two-particle, two-state system. A general two-particle state is a linear combination of these components:

$$g_{00}|00\rangle + g_{01}|01\rangle + g_{10}|10\rangle + g_{11}|11\rangle \qquad (10)$$

where the $g_{ij}$ are numerical coefficients. This state is not necessarily separable. This should be compared with a separable state:

$$|ab\rangle = a_0 b_0|00\rangle + a_0 b_1|01\rangle + a_1 b_0|10\rangle + a_1 b_1|11\rangle \qquad (11)$$

The state (10) is separable if the coefficients of (10) and (11) can be equated. A sufficient condition for separability is

$$g_{00} g_{11} = g_{01} g_{10} \qquad (12)$$

An example of a state which is not separable is,

$$\Phi^+ = \frac{1}{\sqrt{2}}\left(|00\rangle + |11\rangle\right) \qquad (13)$$

The two component states are, however, individually separable. This is, of course, the superposition of two pure two-particle states.

The state $\Phi^+$ is a Bell state and an example of an entangled two-particle state. There is no solution for $a_0, a_1, b_0, b_1$ such that

$$a_0 b_0 = a_1 b_1 = \frac{1}{\sqrt{2}}, \qquad a_0 b_1 = a_1 b_0 = 0 \qquad (14)$$

So the Bell state (13), along with

$$\Phi^- = \frac{1}{\sqrt{2}} \left( |00\rangle - |11\rangle \right)$$

$$\Psi^+ = \frac{1}{\sqrt{2}} \left( |01\rangle \right) + |10\rangle \qquad (15)$$

$$\Psi^- = \frac{1}{\sqrt{2}} \left( |01\rangle - |10\rangle \right)$$

are entangled states. They are, however, clearly linear sums of separable states. Note that for $\Phi^\pm$ the two component one-particle states $|a\rangle$ and $|b\rangle$ have the same state values (either both are 0 or both are 1). For the $\Psi^\pm$ states, the two component one-particle states always have different state values (one is 0 the other is 1). This will be used presently.

A suitable matrix representation for a general two-particle operator is

$$T^{AB} = \begin{bmatrix} T_{00}^A \left[ T^B \right] & T_{01}^A \left[ T^B \right] \\ T_{10}^A \left[ T^B \right] & T_{11}^A \left[ T^B \right] \end{bmatrix}, \qquad \left[ T^B \right] = \begin{bmatrix} T_{00}^B & T_{01}^B \\ T_{10}^B & T_{11}^B \end{bmatrix} \qquad (16)$$

This representation gives an easy example of the well-known (Messiah (1966)) theorem which states that to each operator $T^A$ of the one-particle space $|a\rangle$ there corresponds an operator

$T^{(A)B}$ of the two-particle space $|ab\rangle$ such that if

$$|a'\rangle = T^A |a\rangle \qquad (17)$$

as in (4), then,

$$|a'b\rangle = T^{(A)B} |ab\rangle \qquad (18)$$

    That is, for (18),

$$T^{(A)B} = \begin{bmatrix} T_{00}^A [I] & T_{01}^A [I] \\ T_{10}^A [I] & T_{11}^A [I] \end{bmatrix} = \begin{bmatrix} T_{00}^A & 0 & T_{01}^A & 0 \\ 0 & T_{00}^A & 0 & T_{01}^A \\ T_{10}^A & 0 & T_{11}^A & 0 \\ 0 & T_{10}^A & 0 & T_{11}^A \end{bmatrix}$$

$$[I] = \begin{bmatrix} 1 & 0 \\ 0 & 1 \end{bmatrix} \qquad (19)$$

The two-particle operator which corresponds to the operator

$$T^B = \begin{bmatrix} T_{00}^B & T_{01}^B \\ T_{10}^B & T_{11}^B \end{bmatrix} \qquad (20)$$

is

$$T^{A(B)} = \begin{bmatrix} T^B & 0 \\ 0 & T^B \end{bmatrix} \qquad (21)$$

An example of the correspondence of one-particle transformations to two-particle operators is given by the flip mentioned above.    The two results for the one-particle pure states are

$$F^A|0\rangle = |1\rangle, \qquad F^A|1\rangle = |0\rangle \qquad (22)$$

The corresponding   two-particle operator is

$$F^{(A)B} = \begin{bmatrix} 0 & 0 & 1 & 0 \\ 0 & 0 & 0 & 1 \\ 1 & 0 & 0 & 0 \\ 0 & 1 & 0 & 0 \end{bmatrix} \qquad (23)$$

The action of this operator on the two-particle computational basis is

$$F^{(A)B}|00\rangle = |10\rangle, \quad F^{(A)B}|01\rangle = |11\rangle, \quad F^{(A)B}|10\rangle = |00\rangle, \quad F^{(A)B}|11\rangle = |01\rangle \quad (24)$$

In each case the A particle is flipped while the B particle remains the same.  The flip changes pure states into pure states.

The action on the Bell states is

$$F^{(A)B}\Phi^{\pm} = \Psi^{\pm}, \qquad F^{(A)B}\Psi^{\pm} = \Phi^{\pm} \qquad (25)$$

That is, Bell states with both one-particle component states the same ($\Phi$) transform into Bell states with both one-particle states different ($\Psi$), and vice-versa.

This means that the one-particle transformations produce deterministic changes of the Bell states. This can be contrasted with the probabilistic action of a one-particle transformation on a one-particle state. Thus, these one-particle transformations applied to two-particle states can serve as a means for computation or computer-like operations. This makes an alternative available for quantum computing which is fundamentally different from Feynman computer models which involve multiparticle states. Instead of the parallel processing of multiparticle states, we have sequential processing of one-particle states. Note that the determinism does not violate any of the basic principles of quantum theory. We can establish this by a simple argument:

Refer to equation (25). If we prepare a Bell state, where the component one-particle states have the same value, this state is either $\Phi^+$ or $\Phi^-$, with equal probability. The action of $F^{(A)B}$ on the Bell state produces $\Psi^+$ or $\Psi^-$, which still occur with equal probability. So one of the bits encoded in the Bell state remains completely probabilistic. The second bit, however, is completely determined and can be "switched" as in basic computer operations.

A characterization of the second bit could refer to whether or not the value of the one-particle state A is the "same" as that of B or "different" from B. The flip $F^{(A)B}$ (acting only on A) switches between "same" and "different" producing in turn switching between Bell states $\Phi$ and $\Psi$. Correspondingly the one-particle flip $F^A$ switches between $|0\rangle$ and $|1\rangle$ as in (22).

So far, the measured value of a particular one-particle state has not been considered. The Bell state measurements of same or different do not have anything to say about the values of either of these one-particle states. But these values can, of course, be measured. This will now be studied.

A one-particle state value measurement can be represented at any experimental location by a projection operator. In the simple set of systems being considered here this projection operator is either $|0\rangle\langle 0|$ or $|1\rangle\langle 1|$ for either A or B.

In the case that A is measured to be in the state $|0\rangle$, for the Bell states $\Phi$ (same) this implies that B is in state $|0\rangle$. For the Bell states $\Psi$ (different) this implies that B is in state $|1\rangle$.

A matrix representation of the two-particle projector is

$$P_0^{(A)B} = \begin{bmatrix} 1 & 0 & 0 & 0 \\ 0 & 1 & 0 & 0 \\ 0 & 0 & 0 & 0 \\ 0 & 0 & 0 & 0 \end{bmatrix} = |00\rangle\langle 00| + |01\rangle\langle 01| \qquad (26)$$

The only nonzero value in the matrix (19) is $T_{00}^A = 1$.

But this same projector can be represented as a tensor product of one-particle states,

$$P_0^{(A)B} = |0\rangle\langle 0| \otimes I \qquad (27)$$

When a projector (26) acts on $\Phi^{\pm}$,

$$P_0^{(A)B}\Phi^{\pm} = \frac{1}{\sqrt{2}}\begin{bmatrix} 1 & 0 & 0 & 0 \\ 0 & 1 & 0 & 0 \\ 0 & 0 & 0 & 0 \\ 0 & 0 & 0 & 0 \end{bmatrix}\begin{bmatrix} 1 \\ 0 \\ 0 \\ \pm 1 \end{bmatrix} = \frac{1}{\sqrt{2}}\begin{bmatrix} 1 \\ 0 \\ 0 \\ 0 \end{bmatrix} = \frac{1}{\sqrt{2}}|00\rangle \quad (28)$$

The result is a pure two-particle state.

It can easily be shown that the result of any such projector acting on a Bell state is a pure two-particle state.

So the act of performing a measurement of the value of either of the one-particle states produces a fundamental change in the physics of the experiment. The probabilistic Bell state is changed to a deterministic pure state. There are two different types of experiments, one where the one-particle projector measurement is performed, the other where it is not performed.

The Bell state and the pure state are often discussed as if they were two aspects of the same experiment. There are fundamental differences, however. For example, entanglement is present in the Bell state so long as it exists as a Bell state. There is no entanglement in a pure state. In the two-particle Bell state experiment, when only one qubit is transformed, the two-particle state remains a Bell state. In the pure state experiment, when both qubits are measured the two-particle state remains a pure state. One might ask about experiments where the choice of whether or not to measure is made during the course of the particle through the apparatus. We would say that the results are recorded on the basis of what actually is measured for each particle.

The results are routinely reported as if determined by the Bell state since the experiments are performed for a large number of trials. A test of our viewpoint could be obtained by recording the results of single particle runs.

This would offer an explanation of EPR and the "spooky action-at-a-distance" conundrum. These problems arise when one associates the experimental results of a pure state measurement with the assumption that the state is entangled. When a projection is made to a pure state then the entangled wave function is no longer applicable. Also, when a wave function is that of a Bell state, then it can not apply to an experiment in which the actual value of a one-particle component state is measured.

There are two qubits in the case of a two-particle state. One qubit can be the specification of "same" or "different", the second can be the measured value of one of the single-particle states. Together, these two qubits determine the value of the other single-particle state. Since this is a pure two-particle state, there is no action-at-a distance. The value of the other single-particle state has been determined by the preparation of the experiment.

At first take, this appears to contradict conventional thinking about the probabilistic nature of such experiments. If multiple trials are made with the same apparatus, the results are statistical and seem to indicate that Bell states apply. For example, if the two particles are always in the same state (a $\Phi$ state), then the results are $|00\rangle$ or $|11\rangle$ with equal probability. The average of many trials is just the Bell result. For a measurement of a single run of one experimental set-up, however, the result is either $|00\rangle$ or $|11\rangle$. We claim that this is the deterministic result which would be expected from a pure state. It is only when multiple measurements are made and recorded that the states are reentangled and revert to Bell states. Equivalently, Bell states are produced when two two-particle pure states are superposed (Kwiat et al, 1999). The measurements produce Bell states in the limit of a large number of repetitions.    A single pure state result actually can be compatible with the Bell inequality since the two-particle separable pure state is a product of two one-particle states. The variables are, however, not "hidden" but are observables of the one-particle states. So quantum mechanics is complete "as is" and the existence or absence of hidden variables does not affect its validity. Both Bohr and Einstein were correct. Bohr was right in that Bell states violate the Bell inequality. They cannot be separated into a single product of one-particle states. Einstein was also right since pure states do not violate the Bell inequality. Pure states are separable into a product of one-particle states.

So the foregoing discussion does not conflict with standard treatments of entanglement, Bell states, or quantum measurement. There is, however, a difference in emphasis. Three points are reiterated:

(1) One-particle transformations of components of two-particle Bell states produce corresponding changes in the entangled states. However, Bell states are transformed into Bell states and pure states are transformed into pure states.

(2) The two qubits of these Bell states can be divided into a qubit which refers to whether the two component one-particle states are the same or different and another qubit which gives the measured value of one of the component states.

(3) There is a fundamental dichotomy between experiments in which the value of one of the component states is measured and experiments in which such a value measurement is not made.  A measurement (projector) changes a Bell state into a pure state.

These points lead to a somewhat different rationale for the construction of quantum computers.  All present designs follow the Feynman philosophy which calls for increasing the number of entangled particles as far as is feasible.  This type of processing is massively parallel. In theory, the rationale is attractive, but such problems as decoherence and the necessity of error correction have hindered practical realization.

The proposal here is to exploit the points listed above.  Thus, quantum information would be encoded only in the "same-different" qubit associated with a Bell state.  The information can be manipulated by any number of sequential one-particle relative transformations.  The information is then finally extracted by a relative measurement of the ultimate Bell state.  This is still a relative measurement since the value of the one-particle state is never measured.  (If a projection or measurement of the state value is carried out, then the experiment passes to the realm of pure states.)

A more general example of Bell states is

$$\Phi^{\pm} = s_o|00\rangle \pm \left(1 - s_0^2\right)^{\frac{1}{2}}|11\rangle$$
$$\Psi^{\pm} = \left(1 - s_0^2\right)^{\frac{1}{2}}|01\rangle \pm s_0|10\rangle \qquad (29)$$

For a range of values of $s_0$ from 0 to 1, an array of states could be prepared.  For $s_0=1/\sqrt{2}$ one gets the standard Bell states.  A general measurement of $s_0$ is a Bohr type experiment with a continuous range of state values.

A measurement of same or different ($\Phi$ or $\Psi$) is an Einstein type of experiment with a discrete set of values (two in this case) which are deterministic. The state values can be altered

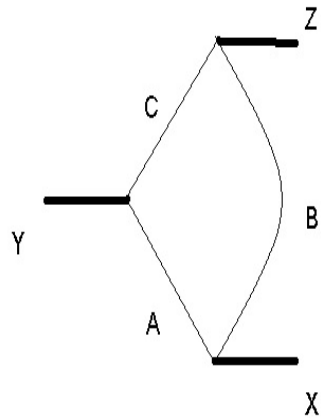

Fig. 1  Quantum logic gate

sequentially by a flip as in (25).  The result (25) holds also for (29)

A simple "circuit" which illustrates this sequential processing method is shown in Fig. 1.

The interaction X produces a pair of systems A and B which have a known relative value for some property.

The interaction Y changes the value of the property of A by a known amount to produce system C.

The interaction Z measures the relative values of A and C.

So the circuit inputs are X and Y, the output is Z.

Again, all preparations and/or measurements are of the relative values of states and not the state values themselves.  These one-particle measurements should be easier to implement and should be more accurate than multiparticle state measurements.

The design of this and similar circuits follows naturally from our viewpoint of the measurement of changes in one-particle systems.  Such circuits would not be likely to appear in current multiparticle design philosophy.  Even so, it is easy to give a correspondence between the one-particle state picture advocated in Fig. 1. and the traditional two-particle state picture.

One can begin with the assumption that X is the pure state |00>.

A general Bell state operator

$$B = \frac{1}{\sqrt{2}} \begin{bmatrix} 1 & 0 & 0 & 1 \\ 0 & 1 & 1 & 0 \\ 0 & 1 & -1 & 0 \\ 1 & 0 & 0 & -1 \end{bmatrix} \qquad (30)$$

accomplishes the entanglement and produces the Bell state $\Phi^+$:

$$B|00> = \Phi^+ \qquad (31)$$

The operator B can be seen to produce each of the four Bell states when it acts on each of the two-particle component (pure) states (9). An example would be a parametric down-conversion.

The interaction Y for a flip, $F^{(A)B}$, produces

$$F^{(A)B}\Phi^+ = \Psi^+ \qquad (32)$$

This could be realized by passing the A photon through a half-wave plate. The operator B again relates the resulting two-particle state

$$B\Psi^+ = |01> \qquad (33)$$

to the pure state | 01>. This is the final value of the two-particle state Z. An example would be second harmonic generation.

Of course, at any point, one *can* make a measurement (e.g. a projection) of a one-particle state value, as in traditional practice. This measurement can disclose the value of another (perhaps distant) state. There is nothing "spooky" about this. Such a measurement mandates that the system passed to a pure state. This is a different experiment, as advocated above. The complicating aspect, then, is that, also in traditional practice, multiple runs are performed and the probabilistic Bell picture reemerges.

In standard Bell discussions (e.g. Nielsen and Chuang (2000), p.117) there are two

assumptions: the assumption of *locality* and the assumption of *realism.* Realism assumes there are physical properties which have definite values. Here, we advocate, not realism, but an assumption which might be called *relative realism,* whereby relative values of the single-particle states within the Bell states can be definite. These relative values, then, are subject to sequential operations or changes.

The argument can go all the way back to Heisenberg, then, and recall that there are no deterministic values for one-particle states. However, it has been shown that there *are* deterministic values for certain properties of two-particle states. These definite or discrete values occur for particular ways of superposing the one-particle states to make two-particle states for example, the Bell states. In such cases there are certain real, definite state values (for example, "same" or "different") which are determined. In simple cases these values may be just the relative values of one component one-particle state to the other. So one might say that the two-particle state has been "decomposed" into an "Einstein bit" which is deterministic and conforms to a relative realism, and a "Bohr bit" which is probabilistic and violates the Bell inequality.

This also applies to two-particle states with more than two possible values. The energy levels of atoms are an example. The relative values of the bound states are determined. The continuum states are probabilistic.

**ACKNOWLEDGEMENTS**

Thanks are in order to  Kenneth L Ketner and Thomas G. McLaughlin for provocative

questions and comments.